\documentclass[twocolumn,showpacs,preprintnumbers,amsmath,amssymb,prl]{revtex4}

\usepackage{graphicx}
\usepackage{dcolumn}
\usepackage{bm}

\usepackage{color}

\begin{document}

\title{Minimal Bending Energies of Bilayer Polyhedra}

\author{Christoph A. Haselwandter and Rob Phillips}

\affiliation{Department of Applied Physics, California Institute of Technology, Pasadena, CA 91125, USA}

\date{\today}

\begin{abstract}
Motivated by recent experiments on bilayer polyhedra composed of amphiphilic
molecules, we study the elastic bending energies of bilayer vesicles forming polyhedral shapes. Allowing for segregation of excess amphiphiles along the ridges of polyhedra, we find that bilayer polyhedra can indeed have lower bending energies than spherical bilayer vesicles. However, our analysis also implies that, contrary to what has been suggested on the basis of experiments, the snub dodecahedron, rather than the icosahedron, generally represents the energetically favorable shape of bilayer polyhedra.
\end{abstract}


\maketitle

In an aqueous environment, amphiphilic molecules such as lipids are observed to self-organize into bilayer vesicles \cite{boal02,safran03}, thus forming the physical basis for cell membranes. Bilayer vesicles generally exhibit
shapes with constant or smoothly varying curvature \cite{seifert97,wortis02}. But in
recent experiments \cite{dubois01,dubois04}, bilayer vesicles with polyhedral shape, consisting of flat faces connected by ridges and vertices with high local curvature, have been observed. In these experiments, two types of oppositely charged, single-tailed amphiphiles were used, with a slight excess of one amphiphile species over the other. At high temperatures, the amphiphiles were found to form spherical bilayer vesicles. However, provided that the number of excess,
unpaired amphiphiles was tuned to some optimal range, cooling the system below the chain melting temperature yielded polyhedral bilayer vesicles.
It was reported that the bilayer polyhedra were stable over weeks, and that their shape was consistently reproduced upon thermal cycling. Furthermore, it was suggested \cite{dubois01,dubois04} that the observed polyhedra had icosahedral symmetry, although some uncertainty regarding the polyhedral
symmetry remained.

What is the mechanism governing the formation and symmetry of bilayer polyhedra? It was argued~\cite{dubois01,dubois04} on the basis of the experimental phenomenology
that elastic contributions to the polyhedron free energy dominate over
entropic or electrostatic contributions, and that minimization of elastic
bending energy alone determines the shape of
bilayer polyhedra. In this Letter, we take these intriguing observations as our starting point
and address the general problem of finding polyhedral shapes with minimal bending energy. Questions regarding the minimal energy shape of bilayer vesicles are commonly answered using a variational approach \cite{seifert97,boal02,safran03}.
However, a given polyhedral shape is defined by the geometric parameters
characterizing its vertices and ridges. Thus, polyhedra are inherently of singular nature, which severely restricts the applicability of variational
calculus. We therefore employ a complementary method, in which we allow for ridges and vertices with
arbitrary geometric properties and, on this basis, calculate polyhedron
bending energies as a function of polyhedron symmetry. In the remainder of this Letter, we first
consider the most straightforward case of bilayers with \textit{uniform}
composition, and then turn to the richer case in which there is segregation
of excess amphiphiles.

The solution of the two-dimensional equations of elasticity is a formidable challenge, and
has only been achieved for the vertices and ridges of polyhedra in certain
limiting
cases \cite{lobkovsky96,lobkovsky97,seung88,lidmar03,nguyen05}. Thus, in order to determine the elastic energies of arbitrary polyhedral shapes, we
mainly employ simple expressions based on the Helfrich-Canham-Evans free energy of bending \cite{boal02,seifert97,safran03},
\begin{equation} \label{helfrich}
G=\frac{K_b}{2} \int dS \left(\frac{1}{R_1}+\frac{1}{R_2}-H_0\right)^2\,,
\end{equation}
where $K_b$ is the bilayer bending rigidity, $R_1$ and $R_2$ are the two principal
radii of curvature, and $H_0$ is the spontaneous curvature. The resulting
expressions for polyhedron energies are intuitive and only involve
a few parameters but, ultimately, are purely phenomenological. We assess their validity by making comparisons to polyhedron energies obtained for the aforementioned limiting cases of the equations of elasticity, which allow for stretching as well as bending deformations.

\begin{figure}[!]
\center
\includegraphics[width=8cm]{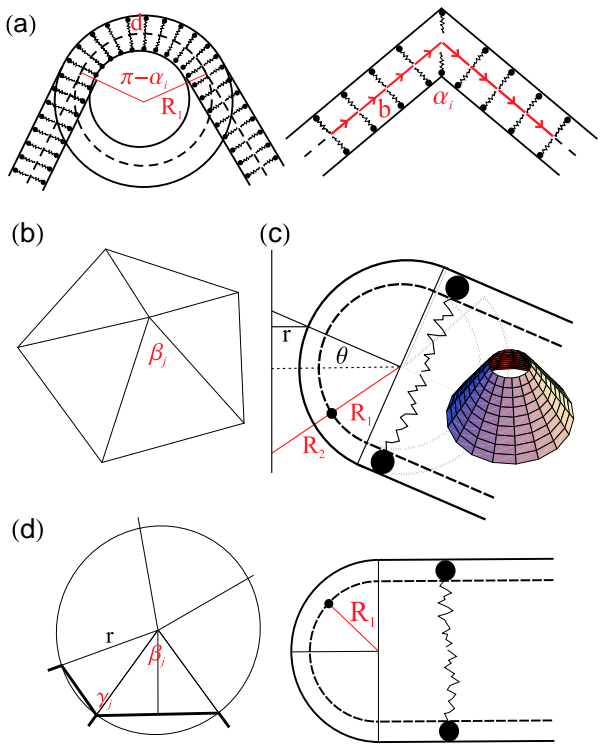}
\caption{\label{fig1}(color online). Illustration of the contributions to the elastic bending
energies of polyhedra: (a) Side view of a ridge with dihedral angle $\alpha_i$, (b) vertex with face angle $\beta_j$, (c) cross section of half of a pore around the tip of a cone (see inset) with apex angle $\pi-2\theta$ and radius $r$, (d) top-down view (left panel) and side view (right panel) of a pore composed of straight edges along each face.}
\end{figure}

Figure~\ref{fig1}(a) shows schematic illustrations of a bilayer bending
gradually (left panel) and sharply (right panel) along a ridge with dihedral
angle $\alpha_i$. The first model is inspired by the electron micrographs of bilayer polyhedra in Refs.~\cite{dubois01,dubois04,glinel04}, while the
second model provides a more faithful representation of the polyhedral geometry.
Based on the picture presented in the left panel of Fig.~\ref{fig1}(a), we approximate ridges by a bilayer bending partially around a cylinder
of radius $R_1=d/(\pi-\alpha_i)$, where $d$ is the arc length. Following the right panel, we discretize
the bilayer with a lattice spacing $b$, and
assume a harmonic potential for the angle between adjacent bond vectors.
Upon setting $d=b$ one finds from either approach a ridge energy similar to the expression used in Ref.~\cite{dubois04},
\begin{equation}  \label{ridgeE}
G_r=\frac{\bar K_b}{2} (\pi-\alpha_i)^2 l\,,
\end{equation} 
where $\bar K_b=K_b/b$, $l$ is the ridge length, and we have assumed that $H_0=0$. A simple expression for the bending energy associated with closed bilayer vertices [see Fig.~\ref{fig1}(b)] is obtained from Eq.~(\ref{helfrich})
following analogous steps, leading to the
vertex energy
\begin{equation} \label{vertexE}
G_v=\frac{K_b}{2} \sum_j \left( \pi- \beta_j \right)^{2}\,,
\end{equation}
where $\beta_j$ denotes the face angle subtended by two ridges meeting at a given polyhedron vertex.

Closed bilayer vertices may break up to form pores \cite{dubois01,dubois04,glinel04},
which was suggested \cite{dubois01} as a mechanism for avoiding the curvature singularity
associated with closed vertices. Figure~\ref{fig1}(c,d) show two models for pores at the vertices of polyhedra
which, similarly as before, are inspired by the experimental images in Refs.~\cite{dubois01,dubois04,glinel04}
[Fig.~\ref{fig1}(c)] and a stricter interpretation of the polyhedral geometry
of bilayer vesicles [Fig.~\ref{fig1}(d)]. In our first model [see Fig.~\ref{fig1}(c)] we approximate the vertex of a given polyhedron by a cone with apex angle $\pi-2\theta$, where $\theta=\pi/2-\arccos \left(1-\Omega/2 \pi \right)$ for a solid angle $\Omega$ subtended by the polyhedron vertex. In our second model we assume that, along each
face, the pore consists of a straight cylindrical edge [see Fig.~\ref{fig1}(d),
left panel], which bends through an angle $\gamma_j$ across a ridge from one face to a neighboring
face [see Fig.~\ref{fig1}(d), right panel]. In both
cases, the elastic pore energy $G_p$ can be evaluated on the basis of Eq.~(\ref{helfrich})
and is found \cite{haselwandter10} to depend on the pore radius, the monolayer bending rigidity, $K_b^\star$, and the monolayer spontaneous curvature, $H_0^\star$,
as well as the face angle and solid angle characterizing
the geometry of a given polyhedron. For physically relevant parameter ranges, $G_p$ increases with decreasing $H_0^\star$ and $\Omega$.

We now turn to the presence of excess amphiphiles in bilayer polyhedra. The expulsion of excess amphiphiles from flat bilayers and the resulting molecular segregation, together with the high spontaneous curvature
of single-tailed excess amphiphiles, are thought \cite{dubois01,dubois04} to have two principal effects on the bending energies of polyhedra. On
the one hand, excess amphiphiles can seed pores into bilayers \cite{dubois01,dubois04} and, thus, pores may have a role
beyond reducing the elastic energy of polyhedron vertices. In particular,
it has been suggested \cite{dubois04} that excess amphiphiles produce pores in the spherical bilayer vesicles from which bilayer polyhedra originate upon
cooling. On the other hand, it has been found~\cite{dubois04} that excess amphiphiles preferentially accumulate along the ridges of polyhedra. As a result, molecular segregation can decrease the bending energy of the outer monolayer at ridges.

The above observations suggest a simple description of how vertex and ridge
energies are modified by the presence of excess amphiphiles.
Ideally, excess amphiphiles are arranged along ridges
such that they induce an anisotropic spontaneous curvature commensurate with the
dihedral angle. Assuming such ``perfect segregation'', only the bending energy of the inner layer
must be considered when computing ridge energies. We therefore obtain a lower bound on the modified ridge energy which takes a similar form as Eq.~(\ref{ridgeE}), but with the rescaled \textit{bilayer} bending modulus $\bar K_b$ replaced by the rescaled \textit{monolayer} bending modulus $\bar K_b^\star=K_b^\star/b$.  Thus, provided that the optimal amount
of excess amphiphiles is present \cite{dubois01,dubois04}, the ridge energy is lowered by a factor $K_b^\star/K_b$, with the number of pores seeded into spherical vesicles equal to or greater than the number of polyhedron vertices. Experiments \cite{dubois01,dubois04} and simulations \cite{hartmann06}
suggest that $K_b^\star/K_b\lessapprox10^{-2}$.

From the simple model for perfectly segregated bilayer polyhedra described
above one finds that the optimal ratio of the amphiphile species in excess to the total amphiphile content is given by $r_I \approx 0.51$ for the polyhedron
sizes observed in experiments \cite{dubois01,dubois04}. This optimal
value for $r_I$ is a direct result of amphiphile and polyhedron geometry and, hence, does not depend on any elastic parameters. The corresponding experimental estimate
is $r_I \approx 0.57$
\cite{dubois01,dubois04}. We expect that in experiments not all excess amphiphiles are segregated
along the ridges and vertices of polyhedra as a result of, for instance, entropic mixing within
bilayer polyhedra or the formation of micelles~\cite{dubois04}. Thus, our theoretical estimate for $r_I$ is in
broad agreement with experimental observations.

For certain limits of the equations of elasticity, approximate solutions corresponding to polyhedron vertex \cite{seung88,lidmar03,nguyen05} and ridge \cite{lobkovsky96,lobkovsky97}
energies have been obtained. In particular, it has been
found \cite{seung88} that 5-fold disclinations in hexagonal lattices are accommodated
for small lattice sizes through a stretching of lattice vectors. However, for
large enough lattice sizes it becomes energetically favorable to buckle out of the plane \cite{seung88}, in which case the energetics of the system are dominated by bending. This behavior is characterized by a dimensionless quantity known as the F\"oppl-von K\'arm\'an number $\Gamma=Y R^2/K_b$, where $Y$ is the two-dimensional Young's modulus and $R$ is the lattice size. As $\Gamma \to \infty$, pronounced ridges develop \cite{lidmar03,nguyen05}
between the 5-fold disclination sites of icosadeltahedral triangulations
of the sphere. In this limit, the total elastic energy is dominated by ridges and was determined in Refs.~\cite{lobkovsky96,lobkovsky97}
to be of the form $K_b (\pi-\alpha)^{7/3} l^{1/3} f(Y/K_b)$, with $0.1\lessapprox f(Y/K_b) \lessapprox
0.4$~nm$^{-1/3}$ for bilayer polyhedra \cite{dubois04,hartmann06,haselwandter10}. Thus, the asymptotic expression for the ridge
energy found in Refs.~\cite{lobkovsky96,lobkovsky97} leads to a similar dependence
on the dihedral angle and proportionality factor as in Eq.~(\ref{ridgeE}), while increasing sub-linearly with
the ridge length $l$.

The lowest energy states of icosadeltahedral
triangulations of the sphere are found to resemble icosahedra for $\Gamma\gtrapprox10^7$
\cite{lidmar03,nguyen05},
which corresponds to a vertex energy greater than $8 K_b$ with, for instance, a value $12K_b$ for $\Gamma=10^{10}$. This compares quite favorably with
the estimate $G_v \approx 11 K_b$ implied by Eq.~(\ref{vertexE}) for the icosahedron. Moreover, we find \cite{haselwandter10} that our two models for pores predict similar ranges for the pore energy $G_p$, with the competition between pores and closed bilayer vertices governed by the ratio $K_b^\star/K_b$. In particular, the aforementioned estimate $K_b^\star/K_b\lessapprox10^{-2}$ \cite{dubois01,dubois04,hartmann06} implies that closed bilayer vertices will be unstable to the formation of (closed) pores, thus removing the singularity associated with polyhedron vertices. This is consistent with experimental
observations \cite{dubois01,dubois04,glinel04} and allows adjustment of the volume of bilayer polyhedra for a fixed total area or number of amphiphiles.

Ridges impose an energetic cost and, hence, one expects that for a fixed area and dihedral angle the faces of polyhedra
relax to form regular polygons. We therefore focus here
on the convex polyhedra with regular polygons as faces, but we have also considered polyhedra with irregular faces \cite{haselwandter10}. The class of convex polyhedra with regular faces encompasses the five Platonic solids, the thirteen Archimedean solids, the two (infinitely large) families of prisms and antiprisms, and the 92 Johnson solids~\cite{cromwell97,berman71}.
It has been shown~\cite{proofJZ} that this list exhausts all convex polyhedra with regular
faces. Thus, counting prisms and antiprisms as one solid each, there
are exactly 112 convex polyhedra with~regular faces. With each of these polyhedra
a specific set of parameters characterizing ridges, vertices, and faces is associated, leading to distinct contributions to the total elastic bending~energy.

\begin{figure}[!]
\center
\includegraphics[width=7.8cm]{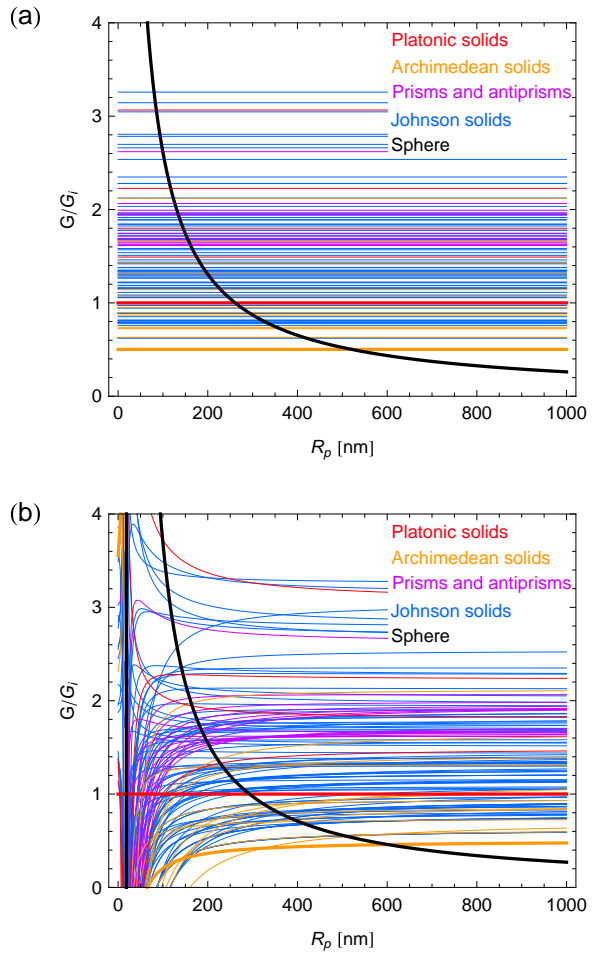}
\caption{\label{fig3}(color online). Total elastic bending energies of the
convex polyhedra with regular faces, obtained for the case of perfect
segregation of excess amphiphiles, and bending energy of the sphere, normalized by the total bending energy of the icosahedron, $G_i$, with (a) pores with $r=0$~nm and (b) pores with $r=20$~nm at each polyhedron vertex. The snub dodecahedron corresponds to the bold curve minimizing bending energy
in (a,b) for $R_p \approx 500$~nm.}
\end{figure}

\begin{figure}[!]
\center
\includegraphics[width=6.4cm]{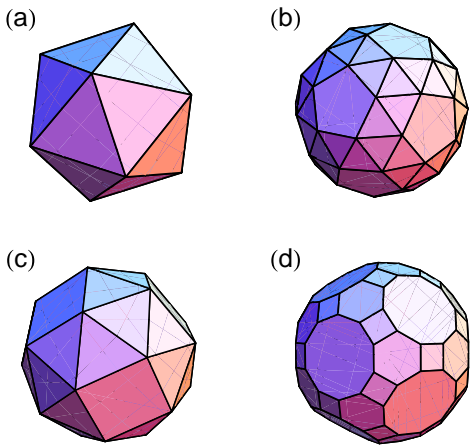}
\caption{\label{fig4}(color online). Image representations of (a)
the icosahedron, (b) the snub
dodecahedron, (c) the snub cube, and (d) the great rhombicosidodecahdron. The polyhedra in (b) and (c) are chiral.}
\end{figure}

We have evaluated the total elastic bending energies of all convex
polyhedra with regular faces. As illustrated in Fig.~\ref{fig3}, polyhedron
energies are compared by plotting elastic energy as a function of the polyhedron radius $R_p$ \cite{lidmar03}, which is related to the polyhedron area $A$ \textit{via} $A=4 \pi R_p^2$, such that, for each value of $R_p$, all shapes have the same total area. While the quantitative details of the resulting
energy curves depend on the particular combination of the aforementioned
expressions for ridge, vertex, and pore energies used, we find that
all curves share the same basic qualitative features. Consistent with a previous
study~\cite{dubois04}, the icosahedron [see Fig.~\ref{fig4}(a)] minimizes bending energy among the Platonic solids. However,
we also find that, in general, the icosahedron does \textit{not} minimize bending
energy among arbitrary polyhedral shapes. In fact, for large enough polyhedron
sizes, the snub dodecahedron [see Fig.~\ref{fig4}(b)] is the polyhedral shape minimizing bending energy among the convex polyhedra with regular faces, and the snub cube [see Fig.~\ref{fig4}(c)] also has a lower energy than the icosahedron in this limit.

Allowing for an optimal number of excess amphiphiles, we find (see Fig.~\ref{fig3}) that polyhedra can
have lower bending energies than the sphere, but only if we permit molecular segregation
along \textit{ridges} as observed in Ref.~\cite{dubois04}. Segregation at pores, which was originally suggested in Ref.~\cite{dubois01} as a potential mechanism stabilizing
polyhedral shapes, is \textit{not} sufficient
to produce polyhedra with bending energies which are favorable compared to the sphere \cite{haselwandter10}. Indeed, if we assume that pores are closed,
bilayer polyhedra are energetically favorable for the experimentally observed polyhedron radius $R_p \approx 500$~nm [see Fig.~\ref{fig3}(a)], with the snub dodecahedron
as the minimum energy shape among the convex polyhedra with regular faces. If we allow pores of a finite size, a sequence of polyhedral shapes is obtained
which minimize bending energy for smaller polyhedron radii [see Fig.~\ref{fig3}(b)].
The most notable of these polyhedral shapes is the great rhombicosidodecahedron
[see Fig.~\ref{fig4}(d)], which surpasses the snub dodecahedron in bending energy at $R_p \approx 300$~nm. However, according to our analysis,
the snub dodecahedron represents the minimum in bending energy among the convex polyhedra with regular faces for the polyhedron sizes and pore sizes ($r \approx 20$~nm) found experimentally \cite{dubois01,dubois04,glinel04}.

In summary, we have used Eq.~(\ref{helfrich}) and a variety
of other expressions~\cite{lobkovsky96,lobkovsky97,seung88,lidmar03,nguyen05} to systematically evaluate the bending energies of bilayer polyhedra. We find that, contrary to what has been suggested on the basis of experiments \cite{dubois01,dubois04}, the snub dodecahedron and the snub cube generally have lower total elastic bending energies than the icosahedron. This result is consistent with several complementary theoretical studies which suggest that the elastic energies of chiral shapes such as the snub dodecahedron and the snub cube can be favorable compared to the icosahedron \cite{bruinsma03,zandi04,dodgson95} and that, even if the icosahedral shape is imposed, the minimum energy structure may still be chiral \cite{vernizzi07}. While we followed here the experimental phenomenology \cite{dubois04,dubois01} and assumed that minimization of bending energy governs the shape of bilayer polyhedra, other contributions to the free energy, as well as kinetic effects \cite{hartmann06,noguchi03}, could,
in principle, modify the preferred polyhedral symmetry.
In light of our results, we suggest revisiting the symmetry of bilayer polyhedra, and the thermodynamic or kinetic mechanisms potentially governing their formation
and stability, in greater experimental detail.

This work was supported by a Collaborative Innovation Award of the Howard
Hughes Medical Institute, and the National Institutes of Health through NIH Award number R01 GM084211 and the Director's Pioneer Award. We thank A. Agrawal,
M. B. Jackson, W. S. Klug, R. W. Pastor, T. R. Powers, D. C. Rees, M. H. B. Stowell, D. P. Tieleman, T. S. Ursell, and H. Yin for helpful comments.

\end{document}